# Timescales of Solar System Formation Based on Al–Ti Isotope Correlation by Supernova Ejecta

**Tsuyoshi Iizuka[1], Yuki Hibiya[2,3], Satoshi Yoshihara[1], Takehito Hayakawa[4]**

[1]Department of Earth and Planetary Science, The University of Tokyo, Hongo 7-3-1, Bunkyo, Tokyo 113-0033, Japan.

[2]Research Center for Advanced Science and Technology, The University of Tokyo, Komaba 4-6-1, Meguro, Tokyo 153-8904, Japan.

[3]Submarine Resources Research Center, Japan Agency for Marine-Earth Science and Technology, Kanagawa 237-0061, Japan.

[4]Quantum Beam Science Research Directorate, National Institutes for Quantum Science and Technology, Umemidai 8-1-7, Kizugawa, Kyoto 619-0215, Japan.

**Abstract**

The radioactive decay of short-lived $^{26}$Al to $^{26}$Mg has been used to estimate the timescales over which $^{26}$Al was produced in a nearby star and the protosolar disk evolved. The chronology commonly assumes that $^{26}$Al was uniformly distributed in the protosolar disk; however, this assumption is challenged by the discordance between the timescales defined by the Al–Mg and assumption-free Pb–Pb chronometers. We find that the $^{26}$Al heterogeneity is correlated with the nucleosynthetic stable Ti isotope variation, which can be ascribed to the non-uniform distribution of ejecta from a core-collapse supernova in the disk. We use the Al–Ti isotope correlation to calibrate variable $^{26}$Al abundances in Al–Mg dating of early solar system processes. The calibrated Al–Mg chronometer indicates a ≥1 Myr gap between parent body accretion ages of carbonaceous and non-carbonaceous chondrites. We further use the Al–Ti isotope correlation to constrain the timing and location of the supernova explosion, indicating that the explosion occurred at 20–30 pc from the protosolar cloud, 0.94 +0.25/–0.21 Myr before the formation of the oldest solar system solids. Our results imply that the Sun was born in association with a ~25 solar mass star.

**Keywords**: Solar system formation (1530); Core-collapse supernovae (304); Meteorites (1038); Protoplanetary disks (1300); Stellar Nucleosynthesis (1616); Isotope shifts (2069)

## 1. Introduction

The short-lived radionuclide (SLR) $^{26}$Al decays to $^{26}$Mg with a half-life ($t_{1/2}$) of 0.73 Myr. Evidence for live $^{26}$Al in the early solar system is provided by $^{26}$Mg variations in meteorites that are correlated with the abundance of a stable isotope $^{27}$Al (Lee et al. 1976). The slopes of the regression lines are equivalent to the initial $^{26}$Al/$^{27}$Al at the time of the formation of the meteorites [($^{26}$Al/$^{27}$Al)$_I$], which reflect their relative formation ages if $^{26}$Al was uniformly distributed in the early solar system. The meteorite Al–Mg ages offer the most precise timeline for events that occurred in the first few Myr of solar system history. Notably, $(^{26}Al/^{27}Al)_I$ values of some components in primitive meteorites (chondrites) are clearly higher than the galactic background level inferred from $\gamma$-ray observations ($^{26}Al/^{27}Al \sim 8 \times 10^{-6}$; Diehl et al. 2006), requiring one or more local stellar sources in/around the parental molecular cloud. However, the nature of the stellar sources remains controversial. Understanding the stellar origin of $^{26}Al$ can constrain the astrophysical setting of solar system formation and the time interval from its nucleosynthesis to the birth of the Sun.

A canonical assumption in the Al–Mg chronology is that the $(^{26}Al/^{27}Al)_I$ of $\sim 5 \times 10^{-5}$ determined for the majority of Ca–Al-rich inclusions (CAIs) in chondrites (Russell et al. 1996; Young et al. 2005; Jacobsen et al. 2008; Larsen et al. 2011) represents that of the solar system as a whole (Villeneuve et al. 2009; Gregory et al. 2020). The CAIs are the oldest known solar system objects which are dated to be $4567.30 \pm 0.16$ Ma by the Pb–Pb chronometer based on the dual decay of $^{235}U–^{207}Pb$ ($t_{1/2} \sim 704$ Myr) and $^{238}U–^{206}Pb$ ($t_{1/2} \sim 4.47$ Gyr) (Amelin et al. 2010; Connelly et al. 2012). It has been demonstrated, however, that some achondrites and chondrules in chondrites have distinctly lower $(^{26}Al/^{27}Al)_I$ than predicted from the canonical value and their Pb–Pb age differences from the CAIs (Schiller et al. 2015; Bollard et al. 2019; Sanborn et al. 2019; Wimipenny et al. 2019) (Fig. 1). This discrepancy provides evidence for non-uniform distribution of $^{26}Al$ in the protosolar disk, unless the U–Pb and Al–Mg systems of the dated objects have closed at different times arising from their distinct behaviors during secondary thermal events (Desch et al. 2023). In addition, bulk samples of carbonaceous chondrites exhibit excesses in the mass-independent abundance of $^{26}Mg$ ($\mu^{26}Mg^*$) compared to those of non-carbonaceous chondrites as well as various achondrites (Larsen et al. 2011; Larsen et al. 2016). Since the parent bodies of carbonaceous chondrites should have accreted at greater heliocentric distances than those of non-carbonaceous meteorites, these observations strongly suggest that [26]Al was more abundant in the outer protosolar disk at a given time. If so, it would call for a revised Al–Mg chronology of solar system formation based on a new approach. In this study, by linking the [26]Al abundance and the nucleosynthetic Ti isotope anomaly in the protosolar disk, we argue their isotope heterogeneity of supernova origin and further validate the use of [26]Al as a chronometer for early solar system events and the nucleosynthesis.

## 2. Al–Ti Isotope Correlation by Supernova Ejecta

Solar system objects show nucleosynthetic stable isotope variations for some non-volatile elements such as Ti, Ni, and Mo (Trinquier et al. 2009; Nanne et al. 2019; Burkhardt et al. 2019). Moreover, the bulk isotope compositions of carbonaceous chondrites are distinct from those of non-carbonaceous chondrites and most achondrites (Warren 2011). The isotope dichotomy reflects the fundamental difference in the contribution of at least one stellar component between the inner and outer protosolar disk. In addition, CAIs have much larger nucleosynthetic anomalies than any bulk meteorites and chondrules, indicating that the stellar component was much more enriched or depleted in the nascent protosolar disk (Brennecka et al. 2020). The uneven distribution of a stellar component in the disk could cause heterogeneity of SLRs if the stellar nucleosynthesis occurred shortly before or during solar system formation. The potential link between the [26]Al heterogeneity and nucleosynthetic stable isotope anomaly has been proposed based on correlated $\mu^{26}$Mg* and $\varepsilon^{54}$Cr variations[4] of primitive and differentiated meteorites (Larsen et al. 2011). Yet, the $\mu^{26}$Mg* variation may be attributed at least partly to nucleosynthetic Mg isotope heterogeneity and Al/Mg variation (Kita et al. 2013). A more rigorous test can be performed using variable $^{26}\text{Al}/^{27}\text{Al}$ at a certain time—for instance at the time of CAI formation [$(^{26}\text{Al}/^{27}\text{Al})_0$]—defined by the meteoritic samples for which $(^{26}\text{Al}/^{27}\text{Al})_I$ are anchored to their absolute Pb–Pb ages (hereafter referred to as time anchors) (Fig. 1).

---

[4] $\varepsilon$ denotes deviation of the mass-bias-corrected isotope ratio from that of a terrestrial standard in parts per 10,000.

Among the time anchors[5], the ungrouped achondrites Northwest Africa (NWA) 2976 (Trinquier et al. 2009) and NWA 6704 (Hibiya et al. 2019; Sanborn et al. 2019) have carbonaceous chondrite-like isotope compositions and, by extension, an outer disk origin, whereas the other achondrites and chondrules have an inner disk origin (Trinquier et al. 2009; Bollard et al. 2019). We find that variable $(^{26}\text{Al}/^{27}\text{Al})_0$ defined by the time anchors are correlated with their $\varepsilon^{50}\text{Ti}$ and $\varepsilon^{46}\text{Ti}$ values (Fig. 2 and Table A1), whereas its correlation with $\varepsilon^{54}\text{Cr}$ values is rather weak mainly due to the data for chondrules whose Ti isotope compositions are unknown. The observed correlation needs to be further verified by establishing more reliable and precise time anchors, but taken at face value, it suggests a common stellar origin for the $^{26}\text{Al}$ heterogeneity and $\varepsilon^{50}\text{Ti}$–$\varepsilon^{46}\text{Ti}$ covariations. The common origin is compatible with the observation that hibonite-rich refractory CAIs formed in the absence of $^{26}\text{Al}$ exhibit extreme $\varepsilon^{50}\text{Ti}$ but few $\varepsilon^{46}\text{Ti}$ anomalies (Kööp et al. 2016).

Our finding provides new insights into the nature of the stellar source of $^{26}\text{Al}$. Asymptotic giant branch (AGB) stars (Wasserburg et al. 2006; Parker & Schoettler 2023), Wolf–Rayet stars (Gaidos et al. 2009; Young 2016; Dwarkadas et al. 2017) as well as their main-sequence progenitors (Gounelle & Meynet 2012), and supernovae (Cameron & Truran 1977; Meyer 2005;

---

[5] Although Erg Chech 002 (EC 002) andesitic meteorite has recently been dated by the Al–Mg (Barrat et al. 2021; Fang et al., 2022; Connelly et al. 2023; Reger et al. 2023) and Pb–Pb methods (Connelly et al. 2023; Krestianinov et al. 2023; Reger et al. 2023), resolvable differences exist in both the reported Al–Mg and Pb–Pb ages, leading to $(^{26}\text{Al}/^{27}\text{Al})_0$ ranging over $1.6$–$4.5 \times 10^{-6}$. Hence, the EC 002 data are not included in the discussion.

Forbes et al. 2021) have been proposed to be potential sources. While stable Ti isotopes can be produced in all these potential stellar sources, the individual sources are characterized by distinct Ti isotope compositions. The correlated $\varepsilon^{50}$Ti–$\varepsilon^{46}$Ti variations among solar system objects have been interpreted to reflect variable mixing of ambient protosolar disk material with an isotopically anomalous component which itself is a homogenized mixture of multiple stellar sources, given that $^{50}$Ti and $^{46}$Ti have different nucleosynthetic origins (Trinquier et al. 2009; Davis et al. 2018). We note, however, that the $\varepsilon^{50}$Ti–$\varepsilon^{46}$Ti covariations are well explained by the heterogeneous contribution of a single stellar component that was synthesized by the weak $s$-process in the C-burning shell of a massive star prior to core collapse (Fig. 2b; see also Appendix A). Although the weak $s$-process is well known to be a major source for light elements ($A < 90$; Käppeler et al. 2011), its significance on the nucleosynthetic anomalies in meteorites has been recognized only by a few studies (Qin et al. 2011; Steele et al. 2012; Nie et al. 2023). Importantly, a weak $s$-process component in the C-rich layers is efficiently ejected by core-collapse supernova (CCSN) explosions when the progenitor stellar mass equals ~25 solar masses ($M_\odot$) (Pignatari et al. 2010); in more massive stars ($\geq 30\ M_\odot$), the weak $s$-process in the C shell becomes less significant relative to that during He core burning in early evolutionary stages, while weak $s$-process products are strongly modified by explosive nucleosynthesis for stars with lower masses (~20 $M_\odot$). On the other hand, $^{26}$Al is significantly produced by the explosive Ne/C-burning in ~25 $M_\odot$ CCSNe (Limongi & Chieffi 2006; Sieverding et al. 2018). Hence, the correlation between $(^{26}$Al/$^{27}$Al$)_0$, $\varepsilon^{50}$Ti, and $\varepsilon^{46}$Ti indicate a nearby ~25 $M_\odot$ CCSN origin for at least a majority of $^{26}$Al. The enrichment of CCSN-ejecta in the outer disk is further supported by recent evidence for heterogeneity of radionuclides $^{92}$Nb (Hibiya et al. 2023) and $^{40}$K (Nie et al. 2023).

## 3. Calibrating the $^{26}$Al Clock for Early Solar System Events

The Al–Ti isotope correlation enables us to calibrate the $^{26}$Al clock for dating early solar system events. Using the non-uniformly distributed $^{26}$Al as an early solar system chronometer requires independent estimates of the variable $(^{26}Al/^{27}Al)_0$ for the source regions of objects used for dating. Given the $(^{26}Al/^{27}Al)_0$–$\varepsilon^{50}$Ti regression line defined by the time anchors (Fig. 2b), we can estimate appropriate $(^{26}Al/^{27}Al)_0$ values for the objects from their $\varepsilon^{50}$Ti. For instance, non-cumulate eucrites, which are considered to represent the basaltic crust of asteroid Vesta, define a whole-rock Al–Mg isochron with an $(^{26}Al/^{27}Al)_t$ of $(4.19 \pm 3.07) \times 10^{-6}$ (Hublet et al. 2017), corresponding to an age of 2.66 +1.39/–0.58 Myr after CAI formation under the assumption of the canonical $(^{26}Al/^{27}Al)_0$ value [$(5.25 \pm 0.02) \times 10^{-5}$; Larsen et al. 2011]. The calibration using the measured $\varepsilon^{50}$Ti of –1.25 ± 0.07 (Supporting data) yields an $(^{26}Al/^{27}Al)_0$ value of $(1.20 \pm 0.20) \times 10^{-5}$ for eucrites, which is identical to the values defined by the angrite time anchors (Schiller et al. 2015) as the result of their Ti isotope similarity (Fig. 2a). This $(^{26}Al/^{27}Al)_0$ value is also indistinguishable from the value of $(1.6 \pm 0.3) \times 10^{-5}$ estimated for the howardite–eucrite–diogenite parent body from the $\varepsilon^{54}$Cr–$\mu^{26}$Mg* relation (Schiller et al. 2011). The calibrated $^{26}$Al clock revises the eucrite age to 1.11 +1.47/–0.59 Myr, rendering Vesta's crust one of the oldest known basaltic crust in the solar system.

The application of the $^{26}$Al clock calibration can be extended to studies of the thermal evolution and accretion timescales of asteroids. Because $^{26}$Al was an important heat source in the early solar system, its variable initial abundance controls the amount of available heat within asteroids at a given early time. Conversely, the initial $^{26}$Al abundance determines the plausible time window for the asteroid accretion to generate internal heat for driving magmatism or metamorphism recorded in meteorites (Hevey & Sanders 2006). Based on the finding that basaltic angrites define a $(^{26}Al/^{27}Al)_0$ value four times lower than the canonical CAI value (Figs. 1 and 2b),

Schiller et al. (2015) showed that the parent body must be accreted within 0.25 Myr after CAI formation so as to allow for large-scale melting, which is 1.4 Myr earlier than the previous estimate. Similar early accretion ages have been inferred for (partially) differentiated asteroids in the inner protosolar disk, assuming either that $^{26}Mg$ deficits in bulk non-carbonaceous meteorites reflect a reduced $(^{26}Al/^{27}Al)_0$ of ~1.6 × 10$^{-5}$ throughout the inner disk (Larsen et al. 2016) or that the asteroids are represented by the ordinary chondrite chondrules defining an $(^{26}Al/^{27}Al)_0$ value of $(1.4 \pm 0.7) \times 10^{-5}$ (Bollard et al. 2019). Furthermore, a delayed accretion at 0.3–1.0 Myr after CAI formation has been proposed for a partially differentiated asteroid in the outer disk using the $(^{26}Al/^{27}Al)_0$ range of 2.7–5.25 × 10$^{-5}$, which was derived from the assumption that the outer disk is a mixture of the inner disk and >10 wt% CAIs (Larsen et al. 2016). In contrast, our calibration method enables a precise estimate of $(^{26}Al/^{27}Al)_0$ for individual meteorites with known Ti isotope compositions. This is significant especially for chondrites because their bulk $(^{26}Al/^{27}Al)_0$ cannot be directly determined as they consist of components with different origins and ages.

The accretion timing of the ordinary and carbonaceous chondrite parent bodies has been estimated to be ~1.8 and 2.1–3.9 Myr after CAI formation to account for the ages and crystallization or closure temperatures of metamorphic minerals (Fujiya et al. 2012; Henke et al. 2012; Fujiya et al. 2013; Doyle et al. 2015; Jogo et al. 2017). The metamorphic ages were determined mainly by the short-lived $^{53}Mn$–$^{53}Cr$ ($t_{1/2}$ ~3.7 Myr) and $^{182}Hf$–$^{182}W$ ($t_{1/2}$ ~8.9 Myr) chronometers, where the SLRs are considered to be homogeneously distributed across the protosolar disk because of the concordance with the Pb–Pb chronometer (Kleine et al. 2012; Sanborn et al. 2019; Tissot et al. 2017) and the homogeneous abundances of the daughter nuclides in a range of bulk chondrites (Trinquier et al. 2008; Kleine et al. 2009). Using ε$^{50}$Ti of bulk ordinary and carbonaceous chondrites (Fig. 2a), we derive $(^{26}Al/^{27}Al)_0$ values ~70 and 40–55% lower than the canonical value, which lead to revised accretion ages of ~0.5 and 1.6–3.1 Myr for the parent bodies, respectively (Fig. 3). These ages are older than the majority of reported chondrule Al–Mg ages (e.g., Pape et al. 2019; Fukuda et al. 2022), but the latter ages assume the canonical $(^{26}Al/^{27}Al)_0$ value and, therefore, are the subject of revision through new Ti isotope analyses of the samples used for dating. On the other hand, our revisions strengthen the case that some chondrules yield Pb–Pb ages younger than the parent body accretion ages (Connelly et al. 2012; Bollard et al. 2017). This observation may reflect either that the young Pb–Pb ages record parent body processes rather than chondrule formation or that the parent bodies did not grow instantaneously but rather gradually and the young chondrules formed and accreted after the major parent body growth. In either case, the revised Al–Mg ages indicate a ≥1 Myr gap between the major accretion epochs of carbonaceous and non-carbonaceous undifferentiated asteroids. This view is in line with numerical models predicting delayed and protracted pebble accretion in the outer disk owing to the outward migration of the snow line (Lichtenberg et al. 2021) (Fig. 4).

## 4. Al–Ti Nuclear Cosmochronology

The initial abundances of SLRs in the solar system have been used to estimate the timing of stellar nucleosynthesis. If a SLR in the early solar system originated exclusively from a nearby single star, its solar initial abundance is depicted as (Wasserburg et al. 2006; Appendix A):

$$\left(\frac{N^{SLR}}{N^{SI}}\right)_0 = d \times \frac{A^{SI} X^{SLR}_{ejecta}}{A^{SLR} X^{SI}_{solar}} \times e^{-\frac{\Delta t}{\tau}}, \tag{1}$$

where $(N^{SLR}/N^{SI})_0$ is the initial abundance ratio of the SLR and a stable isotope (SI) at the time of CAI formation; $d$ is the dilution factor of the stellar ejecta in the solar system; $A^{SLR}$ and $A^{SI}$ are the atomic masses of the SLR and SI; $X^{SLR}_{ejecta}$ and $X^{SI}_{solar}$ are the mass fractions of the SLR and SI in the ejecta and solar system, respectively; $\Delta t$ is the free-decay time between the stellar nucleosynthesis and CAI formation; and $\tau$ is the mean life of the SLR. The optimal values of the two parameters $\Delta t$ and $d$ have been simultaneously obtained to match the solar abundances of multiple SLRs (Looney et al. 2006; Sahijpal & Soni 2006; Takigawa et al. 2008). Yet, the isotope heterogeneity makes it difficult to ascertain the initial SLR abundances of the bulk solar system. On top of that, equation (1) is not applicable to SLRs which have multiple stellar and galactic sources.

Coupling the SLR $^{26}$Al and stable Ti isotope variations offers a new approach to the cosmochronology. Under circumstances where ejecta from a CCSN was non-uniformly distributed in the protosolar disk, the difference in the initial abundance of a CCSN-derived SLR between two disk reservoirs $i$ and $j$ is a function of the free-decay time $\Delta t$ and the difference in the dilution factor between the two reservoirs $\Delta d$ (see Appendix A for more details):

$$\Delta \left( \frac{N^{\mathrm{SLR}}}{N^{\mathrm{SI}}} \right)_{0, i-j} = \Delta d_{i-j} \times \frac{A^{\mathrm{SI}} X_{\mathrm{ejecta}}^{\mathrm{SLR}}}{A^{\mathrm{SLR}} X_{\mathrm{solar}}^{\mathrm{SI}}} \times e^{-\frac{\Delta t}{\tau}} . \qquad (2)$$

Besides, $\Delta d$ can be independently estimated by comparing the meteorite Ti isotope data with the weak $s$-process nucleosynthesis calculations (Fig. 2a), thereby enabling to measure $\Delta t$ by a single SLR chronometer. This new approach has three advantages over the conventional approach. First, it does not require the knowledge of initial SLR abundances of the bulk solar system. Second, it is valid even if the SLR has multiple stellar and galactic sources, as long as the CCSN is responsible for the SLR heterogeneity. Third, by linking the SLR and Ti isotope heterogeneity, the progenitor mass of the CCSN and, therefore, $X_{\mathrm{ejecta}}^{\mathrm{SLR}}$ can be well constrained, which in turn reduces the uncertainty on $\Delta t$ estimate.

Using $\Delta(^{26}\mathrm{Al}/^{27}\mathrm{Al})_0$ of $(4.23 \pm 0.83) \times 10^{-5}$ and $\Delta d$ of $(1.77 \pm 0.15) \times 10^{-3}$ between the CAI and angrite (Sahara 99555) source reservoirs, and applying the predicted $X_{\mathrm{ejecta}}^{\mathrm{Al-26}}$ of $3.73 \times 10^{-6}$ for a 25 $M_\odot$ CCSN from the nucleosynthetic model (Limongi & Chieffi 2006), we obtained $\Delta t = 0.94$

+0.25/–0.21 Myr for the free-decay time between the nucleosynthesis and CAI formation. The CAI formation is thought to occur during the transition of the Sun from a Class I to a Class II protostar (Dauphas & Chaussidon 2011; Brennecka et al. 2020). On the other hand, observations of young stellar objects showed that the typical duration time over a Class 0 and a Class I is ~0.7 Myr (Evans et al. 2009), which is comparable to, or slightly shorter than the free-decay time. Hence, our results suggest that the CCSN occurred during or shortly before the collapse of the protosolar cloud core.

## 5. Implications for Solar System Formation

Two contrasting models of solar system formation have been invoked to explain the nucleosynthetic isotope heterogeneity in the protosolar disk. The first model suggests that the composition of the infalling material from the parental molecular cloud changed from enriched in supernova ejecta in the early stages to depleted in the late stages, and that as a result of viscous expansion of the initial disk, the outer disk contains a higher proportion of early infalling material (Nanne et al. 2019). Such heterogeneous infall is physically feasible especially if CCSN shock wave triggered the collapse of the molecular cloud core; the shock leads to two distinct infall phases, (i) an early infall driven by shock injection and (ii) a late infall when the shock-accelerated protosun traverses more distant and CCSN-ejecta-depleted regions of a molecular cloud (Boss 2022). The second model argues the unmixing of stardust from different stellar sources by thermal processing in the disk, where the inner region is depleted in a thermally labile CCSN component, whereas CAIs reflect the complementary gaseous reservoir rich in this component (Larsen et al. 2011).

Our results have implications for understanding the Sun's birth environment. The dilution factor of the CCSN ejecta in the parental molecular cloud ($d_{cloud}$) can be related to the distance between the CCSN and cloud ($D$) by:

$$D = \frac{r}{2}\sqrt{\frac{\eta}{d_{cloud}} \times \frac{M_{ejecta}}{M_{cloud}}}, \qquad (3)$$

where $M_{ejecta} \sim 23\ M_\odot$ is the total mass of the ejecta, $M_{cloud} \sim 1\ M_\odot$ is the mass of the cloud, $r$ is the radius of the cloud, and $\eta$ is the injection efficiency (Appendix A). In the heterogeneous infall model, CAIs are the most representative objects of the isotope composition of the early collapsing cloud core. In the thermal processing model, CI chondrites would record the chemical and isotope compositions of the parental cloud. The dilution factors for the CAI and CI chondrite source reservoirs ($d_{CAI}$ and $d_{CI}$) can be constrained from the Al–Ti isotope systematics to be within the ranges of $1.6$–$2.4 \times 10^{-3}$ and $0.5$–$1.0 \times 10^{-3}$, respectively (Appendix A). The estimated $d_{CAI}$ range is difficult to reconcile with the CCSN trigger collapse scenario for the protosolar core. Numerical modeling showed that only relatively slow shock waves can trigger core collapse (Boss et al. 2010), requiring $D \sim 10$ pc or more to slow down the shock wave. However, combined with the numerically predicted $\eta$ values of $\leq 0.1$ (Boss & Keiser 2015) and typically observed $r$ values of $\leq 0.1$ pc (Takemura et al. 2023) for cloud cores, the $d_{CAI}$ range yields a $D$ value of $\leq 2$ pc.

Rather, our results are in line with the case where CCSN ejecta were injected into a clumpy molecular cloud prior to core formation (Fig. 4). In this case, injection of CCSN dust would play a more important role in enriching the cloud than gas-phase mixing, resulting in higher $\eta$ values of $\sim 0.4$ (Goodson et al. 2016). Moreover, the dust injection would provide a mechanism for the selective evaporation of dust made from CCSN-derived nuclides in the protosolar disk. The injected dust would be destroyed by sputtering, releasing CCSN-derived nuclides into the cloud gas (Goodson et al. 2016). Subsequently, these nuclides would condense as mantles onto pre-existing dust having different stellar sources. These mantles would be preferentially evaporated during thermal processing in the inner disk (cf., Ek et al. 2019). Using the estimated $d_{CI}$ range and assuming an $r$ value of 0.44 pc for the target clumpy cloud (Goodson et al. 2016), we derive a $D$ value of 20–30 pc. This is comparable to wind-blown bubble sizes of main-sequence stars with initial masses of ~25 $M_\odot$ (Chen et al. 2013), implying that the protosolar molecular cloud was formed by stellar wind compression, and polluted by ejecta from the following CCSN explosion.

## Acknowledgments

This study has become possible thanks to high-precision Al–Mg and Pb–Pb dating and nucleosynthetic isotope characterization of meteoritic samples by multiple laboratories around the world. We acknowledge an anonymous reviewer whose critical and constructive comments helped to improve this paper. We thank Shogo Tachibana, Sota Arakawa, and Kohei Fukuda for discussions. We are grateful to National Institute of Polar Research, Japan for providing the meteorite sample. This work was supported by the Japan Society for the Promotion of Science (grant # 19H01959 to T.I. and Y.H.; grant #22H00170 to T.I. and T.H.).

## Appendix A

### Cosmochronology using Isotope Heterogeneity

Short-lived radionuclides (SLRs) present in the early solar system were either inherited from the galactic background or injected by nearby stars into the molecular cloud from which the Sun formed (Wasserburg et al. 2006; Huss et al. 2009). The initial abundances of SLRs in the solar system have been used to determine the time interval between the isolation of the parental molecular cloud from uniform SLR production in the galaxy and the formation of the solar system, assuming that the late SLR addition by local stars is insignificant (Lugaro et al. 2014).

Alternatively, in a scenario based on which a significant fraction of SLRs was injected into the solar parental molecular cloud by a nearby star, the initial SLR abundances are functions of two parameters: the time interval from the stellar nucleosynthesis to the formation of the solar system and the dilution factor of the injected stellar ejecta relative to the target molecular cloud. The optimal values of the two parameters have been estimated to match the initial abundances of multiple SLRs (Looney et al. 2006; Sahijpal & Soni 2006; Takigawa et al. 2008). In either case, the use of SLRs as cosmochronometers requires the knowledge of their initial abundances in the bulk solar system. However, CCSN-derived $^{26}$Al was heterogeneously distributed in the protosolar disk, making it difficult to ascertain their initial abundances in the bulk solar system. We present a cosmochronology method using a SLR that was non-uniformly distributed in the protosolar disk.

We tentatively consider that the SLR originated solely from a nearby star, without a galactic inventory. Let $N_{\text{ejecta}(i)}^{\text{SLR}}$ and $N_{\text{ejecta}(i)}^{\text{SI}}$ be the numbers of SLR and the stable isotope (SI) in stellar ejecta that contaminated reservoir $i$ in the protosolar disk, respectively; $N_{\text{disk}(i)}^{\text{SI}}$ be the number of the SI in the ambient disk without the contamination; and $(N^{\text{SLR}}/N^{\text{SI}})_{0,i}$ be the abundance ratio of the SLR and SI in the reservoir at the time of the formation of CAIs, we obtain:

$$\left(\frac{N^{\text{SLR}}}{N^{\text{SI}}}\right)_{0,i} = \frac{N_{\text{ejecta}(i)}^{\text{SLR}} \times e^{-\frac{\Delta t}{\tau}}}{N_{\text{disk}(i)}^{\text{SI}} + N_{\text{ejecta}(i)}^{\text{SI}}}, \tag{A1}$$

where $\Delta t$ is the free-decay time interval between the SLR nucleosynthesis and CAI formation and $\tau$ is the mean life of the SLR. The parameter $\Delta t$ becomes negative when the SLR was synthesized after CAI formation and directly injected into the protosolar disk. By introducing the dilution factor for the reservoir $d_i$, that is, the mass ratio of the incorporated ejecta to the ambient disk material, we can rewrite equation (A1) as follows:

$$\left(\frac{N^{\text{SLR}}}{N^{\text{SI}}}\right)_{0,i} = \frac{d_i \times Q_{\text{ejecta}}^{\text{SLR}} \times e^{-\frac{\Delta t}{\tau}}}{Q_{\text{disk}}^{\text{SI}} + d_i \times Q_{\text{ejecta}}^{\text{SI}}}, \tag{A2}$$

where $Q_{\mathrm{disk}}^{\mathrm{SI}}$ is the number of the SI per unit mass of the ambient disk and $Q_{\mathrm{ejecta}}^{\mathrm{SLR}}$ and $Q_{\mathrm{ejecta}}^{\mathrm{SI}}$ are the numbers of the SLR and SI per unit mass of the incorporated ejecta, respectively. If $d_i \ll 1$, we can approximate that the denominator term of the right side is constant among different reservoirs in the disk. Thus, we obtain:

$$\left(\frac{N^{\mathrm{SLR}}}{N^{\mathrm{SI}}}\right)_{0,i} \approx \frac{d_i \times Q_{\mathrm{ejecta}}^{\mathrm{SLR}} \times e^{-\frac{\Delta t}{\tau}}}{Q_{\mathrm{solar}}^{\mathrm{SI}}}, \tag{A3}$$

where $Q_{\mathrm{solar}}^{\mathrm{SI}}$ is the number of SI per unit mass of the solar system. The difference in the isotope ratio between reservoirs $i$ and $j$ in the disk is related to the difference in their dilution factors:

$$\Delta\left(\frac{N_0^{\mathrm{SLR}}}{N_0^{\mathrm{SI}}}\right)_{i-j} = \Delta d_{i-j} \times \frac{Q_{\mathrm{ejecta}}^{\mathrm{SLR}}}{Q_{\mathrm{solar}}^{\mathrm{SI}}} \times e^{-\frac{\Delta t}{\tau}}. \tag{A4}$$

Importantly, this relation holds even if galactic and other stellar inventories of the SLR do exist. Because stellar nucleosynthesis yields are generally expressed in solar mass units, we may rewrite equations (A3) and (A4) to equations (1) and (2) using the mass fraction of the SLR in the incorporated ejecta ($X_{\mathrm{ejecta}}^{\mathrm{SLR}}$), the mass fraction of the SI in the solar system ($X_{\mathrm{solar}}^{\mathrm{SI}}$), and atomic masses of the SLR and SI ($A^{\mathrm{SLR}}$ and $A^{\mathrm{SI}}$).

The parameters $\Delta t$ and $\Delta d$ can be uniquely defined by combining multiple SLR cosmochronometers with the same stellar origin. Alternatively, $\Delta d$ can be independently estimated from the variation in the nucleosynthetic stable isotopes, which in turn allows for the measurement of $\Delta t$ using a single SLR cosmochronometer. In practice, however, the extent of isotope heterogeneity in the disk may differ depending on elements, which results in a difference in $\Delta d$ between multiple isotope systematics. Indeed, variations in nucleosynthetic stable isotopes are evident in bulk meteorites with respect to refractory elements such as Ti (Trinquier et al. 2009) and Mo (Budde et al. 2016; Brennecka et al. 2020), but absent for many volatile elements such as Cd (Wombacher et al. 2008; Toth et al. 2020) and Te (Fehr et al. 2005). Although Pd, that is, a moderately refractory element, displays nucleosynthetic anomalies, the magnitudes are smaller than those predicted based on more refractory elements in the same mass region (Ek et al. 2020). These observations indicate that the elemental condensation temperature controls the variation of dilution factor in the disk, likely due to the selective evaporation of a CCSN-component in the inner disk in concert with outward transport of CAIs condensed from the gas enriched in this component (Fig. 4). The significance of CAI recycling into the outer disk is supported by the element patterns of chondrites (Fig. A1), showing that CAI-bearing carbonaceous chondrites are relatively depleted in elements with 50% condensation temperatures ($T_{50}$) below 1350 K as compared to near CAI-free ordinary, enstatite, and CI chondrites (van Kooten et al. 2024).

In this study, we used the meteorite Ti isotope data to estimate $\Delta d$ and then applied it to the $^{26}$Al cosmochronometer. Because both Al and Ti are refractory elements with a $T_{50}$ above 1550 K, they are expected to have a common $\Delta d$ value. The application can potentially be extended to other SLRs, provided they have a CCSN origin and high condensation temperature (i.e., $T_{50} > 1350$ K). By contrast, volatile SLRs, such as $^{129}$I, would have been evenly distributed in the early solar system and are thus suitable for the conventional cosmochronology.

### A.1. Nucleosynthetic Ti Isotope Variation

Solar system objects exhibit significant variations in $\varepsilon^{50}$Ti and $\varepsilon^{46}$Ti, which correlate with each other (Fig. 2a and Supporting data). The measured Ti isotope ratios are internally normalized by fixing their $^{49}$Ti/$^{47}$Ti ratios to the terrestrial value of 0.749766 (Niederer et al. 1981), but the $^{49}$Ti/$^{47}$Ti ratios are actually variable due to nucleosynthetic anomalies. Thus, the $\varepsilon^{50}$Ti–$\varepsilon^{46}$Ti covariations reflect differential nucleosynthetic anomalies of $^{50}$Ti–$^{47}$Ti–$^{49}$Ti and $^{46}$Ti–$^{47}$Ti–$^{49}$Ti, respectively. While these Ti isotopes can be produced in $s$-processes in AGB and massive stars as well as explosive processes in supernovae, the individual processes are characterized by distinct

Ti isotope compositions. Hence, the notable $\varepsilon^{50}$Ti–$\varepsilon^{46}$Ti correlation is an important clue to the stellar source responsible for the nucleosynthetic isotope variations.

Let us consider that Ti isotope variations are caused by heterogeneous incorporation of stellar ejecta into the protosolar disk. Based on equation (A1), the abundance ratio of a Ti isotope, $^x$Ti, to $^{47}$Ti for reservoir $i$ in the disk is given by:

$$\left(\frac{^x\text{Ti}}{^{47}\text{Ti}}\right)_i = \frac{Q_{\text{disk}}^{\text{Ti}-x} + d_i \times Q_{\text{ejecta}}^{\text{Ti}-x}}{Q_{\text{disk}}^{\text{Ti}-47} + d_i \times Q_{\text{ejecta}}^{\text{Ti}-47}} = \frac{Q_i^{\text{Ti}-x}}{Q_i^{\text{Ti}-47}}. \tag{A5}$$

The isotope ratio in the reservoir relative to that in another reservoir $j$ is expressed by:

$$\left(\frac{^x\text{Ti}}{^{47}\text{Ti}}\right)_i \Big/ \left(\frac{^x\text{Ti}}{^{47}\text{Ti}}\right)_j = \frac{1 + \Delta d_{i-j} \times Q_{\text{ejecta}}^{\text{Ti}-x}/Q_j^{\text{Ti}-x}}{1 + \Delta d_{i-j} \times Q_{\text{ejecta}}^{\text{Ti}-47}/Q_j^{\text{Ti}-47}} \approx \frac{1 + \Delta d_{i-j} \times Q_{\text{ejecta}}^{\text{Ti}-x}/Q_{\text{solar}}^{\text{Ti}-x}}{1 + \Delta d_{i-j} \times Q_{\text{ejecta}}^{\text{Ti}-47}/Q_{\text{solar}}^{\text{Ti}-47}}. \tag{A6}$$

Using the mass fractions of the nuclides $X$ instead of their numbers per unit mass $Q$, this can be rewritten as:

$$\left(\frac{^x\text{Ti}}{^{47}\text{Ti}}\right)_i \Big/ \left(\frac{^x\text{Ti}}{^{47}\text{Ti}}\right)_j \approx \frac{1 + \Delta d_{i-j} \times X_{\text{ejecta}}^{\text{Ti}-x}/X_{\text{solar}}^{\text{Ti}-x}}{1 + \Delta d_{i-j} \times X_{\text{ejecta}}^{\text{Ti}-47}/X_{\text{solar}}^{\text{Ti}-47}}. \tag{A7}$$

Because Ti isotope ratios measured on solar system objects are internally normalized to the fixed $^{49}$Ti/$^{47}$Ti ratio with the exponential law, we obtain:

$$\left(\frac{^x\text{Ti}}{^{47}\text{Ti}}\right)_i^* = \left(\frac{^x\text{Ti}}{^{47}\text{Ti}}\right)_i \times \left(\frac{A^{\text{Ti}-x}}{A^{\text{Ti}-47}}\right)^{-\beta}, \tag{A8}$$

$$\beta = ln\left[\left(\frac{^{49}\text{Ti}}{^{47}\text{Ti}}\right)_i \Big/ 0.749766\right] \Big/ ln\left(\frac{A^{\text{Ti}-49}}{A^{\text{Ti}-47}}\right), \tag{A9}$$

where the asterisk denotes the internally normalized ratio. Hence, normalized Ti isotope ratios in the different reservoirs are related as follows:

$$\left(\frac{^x\text{Ti}}{^{47}\text{Ti}}\right)_i^* \Big/ \left(\frac{^x\text{Ti}}{^{47}\text{Ti}}\right)_j^* \approx \frac{1 + \Delta d_{i-j} \times X_{\text{ejecta}}^{\text{Ti}-x}/X_{\text{solar}}^{\text{Ti}-x}}{1 + \Delta d_{i-j} \times X_{\text{ejecta}}^{\text{Ti}-47}/X_{\text{solar}}^{\text{Ti}-47}} \times \left(\frac{A^{\text{Ti}-x}}{A^{\text{Ti}-47}}\right)^{ln\left(\frac{1 + \Delta d_{i-j} \times X_{\text{ejecta}}^{\text{Ti}-47}/X_{\text{solar}}^{\text{Ti}-47}}{1 + \Delta d_{i-j} \times X_{\text{ejecta}}^{\text{Ti}-49}/X_{\text{solar}}^{\text{Ti}-49}}\right) \Big/ ln\left(\frac{A^{\text{Ti}-49}}{A^{\text{Ti}-47}}\right)}. \tag{A10}$$

The mass fractions of the Ti isotopes in the solar system ($X_{solar}^{Ti-x}$) were taken from Lodders (2003). The $X_{ejecta}^{Ti-x}$ was calculated for the potential stellar sources using Ti isotope yields reported for explosive processes in a SNIa [O-DDT model in Maeda et al. 2010] and a 15 $M_\odot$ CCSN [S15 model in Rauscher et al. 2002], the main s-process in a 3 $M_\odot$ AGB star [m3z1m2 model in Battino et al. 2019], and weak s-processes in the He-burning core of a 30 $M_\odot$ star (The et al. 2010) and in C-burning shell of a 25 $M_\odot$ star [Model 1 in Pignatari et al. 2010] (Table A2). With respect to the $X_{ejecta}^{Ti-x}$ calculations of the weak s-process components, we assume that the components synthesized during pre-CCSN stages were ejected by a CCSN explosion without significant modification and that the central 2 $M_\odot$ part of the stars fell back onto a collapsing stellar core (remnant mass). The $\varepsilon^{50}$Ti–$\varepsilon^{46}$Ti trends produced by admixing the individual stellar components to the angrite source reservoir were compared with the meteorite data (Fig. 2a), showing that the mixing trend with the weak s-process component in the C-burning shell agrees well with the $\varepsilon^{50}$Ti–$\varepsilon^{46}$Ti covariations recorded by the meteoritic samples. This agreement is consistent with the Al–Ti isotope correlation (Fig. 2b) because the weak s-process component synthesized in C-rich layers and $^{26}$Al synthesized in C- and Ne-rich layers in the early phase of CCSN are ejected by a single explosion event. While the addition of the weak s-process component is expected to cause a deficit in $\varepsilon^{48}$Ti, most solar system objects show no significant $\varepsilon^{48}$Ti variation. Considering that $^{48}$Ti can be preferentially synthesized by the $\alpha$-rich freeze-out in the inner regions of CCSNe, the lack of $\varepsilon^{48}$Ti deficits may reflect that the CCSN ejected a small fraction of inner material exposed to the $\alpha$-rich freeze-out with a mixing fallback mechanism (e.g., Nomoto et al. 2006). In addition, the enrichment of the weak s-process component in the outer disk may account for $\varepsilon^{48}$Ca and $\varepsilon^{54}$Cr excesses in carbonaceous meteorites (Trinquier et al. 2007; Qin et al. 2011; Schiller et al. 2018). This possibility needs to be further explored by quantifying the contribution of the $\alpha$-rich freeze-out component for Ca and the condensation temperature effect on Cr isotope variations ($T_{50}$ = 1296 K) in the protosolar disk, which is beyond the scope of this study.

The comparison of the meteorite data with the mixing line defined by the weak $s$-process component (Fig. 2a) further allows us to estimate the changes in the dilution factor of the CCSN ejecta among the meteorite source reservoirs. The CAIs analyzed for Al–Mg isotopes display excesses in $\varepsilon^{50}$Ti and $\varepsilon^{46}$Ti of 10.59 ± 1.13 and 1.90 ± 0.16, respectively, relative to the angrite average (Table A1), which are comparable to $\Delta d$ values of (1.84 ± 0.20) × 10⁻³ and (1.77 ± 0.15) × 10⁻³, respectively. Note that while the angrite source reservoir could be contaminated by the CCSN ejecta to some extent (i.e., $d_{\mathrm{angrite}} \geq 0$), the $\Delta d$ estimate is insensitive to the extent of the contamination.

### A.2. Aluminum-26 Cosmochronometer

The $(^{26}\mathrm{Al}/^{27}\mathrm{Al})_0$ defined by CAIs is (4.23 ± 0.83) × 10⁻⁵ higher than that by the Sahara 99555 angrite (Table A1). Combining this $(^{26}\mathrm{Al}/^{27}\mathrm{Al})_0$ difference with the $\Delta d$ value of (1.77 ± 0.15) × 10⁻³ estimated from the Ti isotope systematics allows to estimate the timing of the CCSN explosion responsible for the isotope heterogeneity. For the $^{26}$Al cosmochronometer, we can rewrite equation (2) as:

$$\Delta \left( \frac{^{26}\mathrm{Al}}{^{27}\mathrm{Al}} \right)_{0,\mathrm{CAI-angrite}} = \Delta d_{\mathrm{CAI-angrite}} \times \frac{A^{\mathrm{Al-27}} X_{\mathrm{ejecta}}^{\mathrm{Al-26}}}{A^{\mathrm{Al-26}} X_{\mathrm{solar}}^{\mathrm{Al-27}}} \times e^{-\frac{\Delta t}{\tau}}, \qquad (A11)$$

where $X_{\mathrm{solar}}^{\mathrm{Al-27}}$ = 6.64 × 10⁻⁵ (Lodders 2003) and $\tau$ = 1.05 Myr. The $X_{\mathrm{ejecta}}^{\mathrm{Al-26}}$ was calculated from a $^{26}$Al yield of 8.57 × 10⁻⁵ $M_\odot$ predicted for a 25 $M_\odot$ CCSN with a remnant mass of 2 $M_\odot$ (Limongi & Chieffi 2006) because such CCSN can yield the weak $s$-process component generated in the C-burning shell (Pignatari et al. 2010) (Table A2). Using the calculated $X_{\mathrm{ejecta}}^{\mathrm{Al-26}}$ value of 3.73 × 10⁻⁶, we derive a free-decay time $\Delta t$ of 0.94 +0.25/–0.21 Myr.

### *A.3. Distance between the Supernova and Protosolar Cloud*

Assuming a spherically symmetric ejection from the CCSN, the distance $D$ between the protosolar molecular cloud and CCSN is related to $f$, that is, the ratio of the amount of a nuclide injected into the cloud to the total amount of the nuclide ejected from the CCSN:

$$f = \frac{\pi r^2}{4\pi D^2} \times \eta, \tag{A12}$$

where $r$ is the radius of the cloud and $\eta$ is the injection efficiency, which accounts for the fact that only a part of the incoming ejecta may be trapped by the cloud. Numerical modeling (Boss & Keiser 2015) indicated that in the triggered collapse scenario, CCSN material is injected into the collapsing core essentially by gas-phase mixing through Rayleigh–Taylor instabilities with $\eta$ values of ~0.04–0.1. On the other hand, Goodson et al. (2016) modeled the interaction between CCSN ejecta dust and clumpy molecular cloud prior to core formation, and predicted $\eta$ values of ~0.4. Note that while CCSN ejecta dust can potentially be injected into an already-formed protoplanetary disk (Ouellette et al. 2010), this scenario is inconsistent with the free-decay time obtained in this study. Note also that $f$ (or its inverse) is called "dilution factor" in some literature. When all nuclides are injected into the cloud with the same efficiency, $f$ can be converted to our dilution factor as follows:

$$d_{\text{cloud}} = f \times \frac{M_{\text{ejecta}}}{M_{\text{cloud}}}, \tag{A13}$$

where $M_{\text{ejecta}}$ is the total mass of the ejecta and $M_{\text{cloud}}$ is the mass of the parental cloud. Accordingly, the relation between $D$ and $d_{\text{cloud}}$ can be expressed by equation (3).

Among the solar system objects, CAIs are the oldest solar system solids and are enriched the most in the CCSN component. In the framework of the heterogeneous infall model (Nanne et al. 2019), therefore, CAIs are the best representative of the isotope composition of an early collapsing cloud core that has been subjected to the injection of CCSN ejecta. In the context of the thermal processing model (Larsen et al. 2011), CI chondrites, whose chemical compositions are closest to that of the solar photosphere, are considered as representative samples of the parental cloud.

The dilution factors for the CAI and CI chondrite source reservoirs ($d_{CAI}$ and $d_{CI}$) cannot be explicitly defined without knowledge of the nucleosynthetic contribution of the CCSN causing the isotope heterogeneity relative to those of other stellar and galactic sources. Nevertheless, the $\Delta d_{CAI\text{-}angrite}$ value of $(1.77 \pm 0.15) \times 10^{-3}$ estimated from the Ti isotope data provides a lower limit for the $d_{CAI}$. Moreover, the upper limit of the $d_{CAI}$ is derived by considering the extreme case in which all $^{26}$Al in the early solar system originated from the CCSN. In this case, the $d_{CAI}$ must be a factor of $5.15 \pm 1.43$ higher than $d_{angrite}$, which, combined with the $\Delta d_{CAI\text{-}angrite}$ value, leads to a $d_{CAI}$ value of $(2.20 \pm 0.24) \times 10^{-3}$. Consequently, the $d_{CAI}$ value is constrained to be within the range of 1.6–$2.4 \times 10^{-3}$. In the same manner, the range of allowable $d_{CI}$ is estimated to be 0.5–$1.0 \times 10^{-3}$. If the CAI source reservoir was a mixture of the early collapsing cloud core and the late infalling cloud envelope, the dilution factor for the core would be larger than $d_{CAI}$, resulting in a smaller $D$ value.

## Appendix B

### Titanium Isotope Analysis of Asuka 881394

In this study, we determined the Ti isotope composition of Asuka 881394 for which Ti isotope data were not reported previously. A ~0.15 g interior chip of the ungrouped basaltic meteorite Asuka 881394 was crushed in an agate mortar. Two ~50 mg aliquots of the powdered sample were digested with a concentrated HF–HNO$_3$ mixture at 200 ˚C in a 125 mL Parr® bomb, followed by repeated evaporation with concentrated HNO$_3$ and dissolution in 6 M HCl. The aliquots were processed through a three-step column chemistry. In the first step, the sample in 6 M HCl was loaded onto the column packed with Bio-Rad AG1-X8 anion exchange resin (200-400 mesh), in which Ti is eluted whereas Fe and U are retained by the resin. The second step utilizes Eichrom

DGA resin (50–100 μm mesh). In this step, matrix elements including Cr and Ca were extracted in 12 M $HNO_3$ followed by Ti elution in 12 M $HNO_3$ + 1wt% $H_2O_2$. Finally, Ti was further purified using AG1-X8 resin, where residual matrix elements were extracted in 4 M HF and 0.4 M HCl + 1 M HF, followed by Ti elution in 1 M HCl + 2% $H_2O_2$.

The Ti isotope ratio measurements were performed on a Neptune plus multiple collector-inductively coupled plasma mass spectrometer (Thermo Fisher Scientific) attached to an Aridus II micro-concentric desolvating nebulizer (Cetac Technology). The separated Ti fraction was dried and dissolved in 0.5 M $HNO_3$ containing a trace amount of HF to a concentration of ~200 ng/mL. Measurements were carried out using a Jet sample cone and a X skimmer cone with medium mass resolution and a sample uptake rate of ~0.1 mL/min, which resulted in $^{48}$Ti signal intensities of 3.0–3.5 × 10$^{-10}$ A. All five Ti isotopes together with $^{43}$Ca, $^{51}$V, and $^{53}$Cr were monitored in dynamic mode using 9 Faraday cups. Data were acquired from 40 cycles, 2 lines/cycle, 8.4 s integration/line, and 3 s idle time between lines. Instrumental mass fractionation was corrected relative to $^{49}$Ti/$^{47}$Ti = 0.749766 (Niederer et al. 1981) with an exponential law. For the isobaric interference correction, the mass bias factors for Ca, V, and Cr were assumed to be identical to that for Ti, and the literature values of $^{50}$V/$^{51}$V = 0.0251 and $^{50}$Cr/$^{53}$Cr = 0.4573 (Rosman et al. 1998) and empirically derived ratios of $^{48}$Ca/$^{43}$Ca = 0.0237 and $^{46}$Ca/$^{43}$Ca = 1.4039 were utilized. This interference correction produces accurate results for Ti standard solutions doped with Ca, V, and Cr at abundances more than one order of magnitude higher than the analyzed samples ($^{43}$Ca/$^{47}$Ti ≤ 3 × 10$^{-4}$, $^{51}$V/$^{47}$Ti ≤ 1 × 10$^{-5}$, and $^{53}$Cr/$^{47}$Ti ≤ 5 × 10$^{-5}$). Individual sample measurements were bracketed by analyses of an Alfa Aesar Ti standard solution. The Ti isotope ratios in samples were determined relative to the standard runs. The two sample aliquots were analyzed three times each (Table B1) and yielded mean ± 95% confidence interval values of $\varepsilon^{46}Ti = -0.19 \pm 0.05$, $\varepsilon^{48}Ti = -0.06 \pm 0.05$, and $\varepsilon^{50}Ti = -1.14 \pm 0.09$.

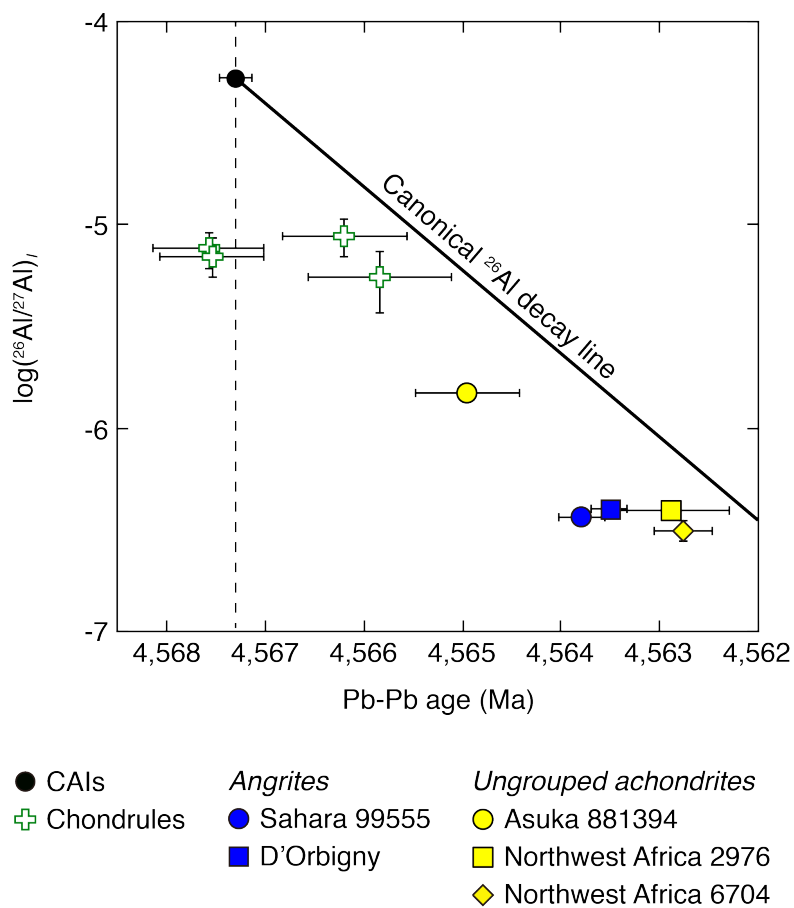

**Figure 1. ($^{26}$Al/$^{27}$Al)$_I$ versus Pb–Pb age plots for the meteoritic samples.** The solid line represents the canonical $^{26}$Al decay line defined by CAIs (Larsen et al. 2011). The vertical dashed line indicates the time of CAI formation (4567.30 Ma; Amelin et al. 2010; Connelly et al. 2012). All the Pb–Pb ages were calculated using the measured $^{238}$U/$^{235}$U of the individual samples. Error bars represent 2σ. The data are listed in Table A1.

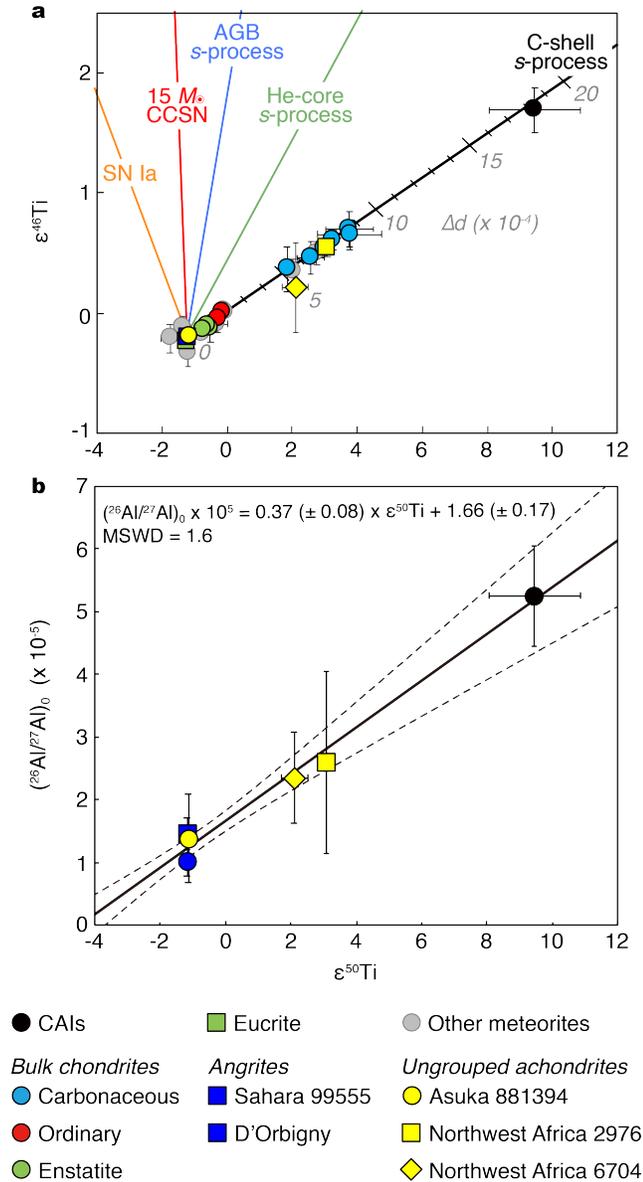

**Figure 2.** Titanium isotopes as a proxy for the nucleosynthetic heterogeneity. (a) $\varepsilon^{46}$Ti–$\varepsilon^{50}$Ti variation diagram for solar system objects. Mixing trends between the angrite source reservoir and five potential stellar Ti sources are shown for comparison: post-explosion compositions of SN Ia (Maeda et al. 2010) and 15 $M_\odot$ CCSN (Rauscher et al. 2002), an AGB $s$-process component (Battino et al. 2019), and weak $s$-process components in the He-burning core (The et al. 2000) and C-burning shell (Pignatari et al. 2010) in massive stars. The ticks on the mixing line with the C shell component denote the dilution factor $\Delta d$ of the ejecta at $10^{-4}$ intervals. The Ti isotope data are listed in Supporting data. (b) $(^{26}$Al/$^{27}$Al$)_0$ versus $\varepsilon^{50}$Ti for the meteoritic samples. The $(^{26}$Al/$^{27}$Al$)_0$ values were obtained by projecting $(^{26}$Al/$^{27}$Al$)_t$ of the individual meteorites onto the time of CAI formation at 4567.30 Ma using their Pb–Pb ages and the decay rate of $^{26}$Al (Table A1). Solid and dashed lines represent the regression and 95% confidence intervals of the data, respectively.

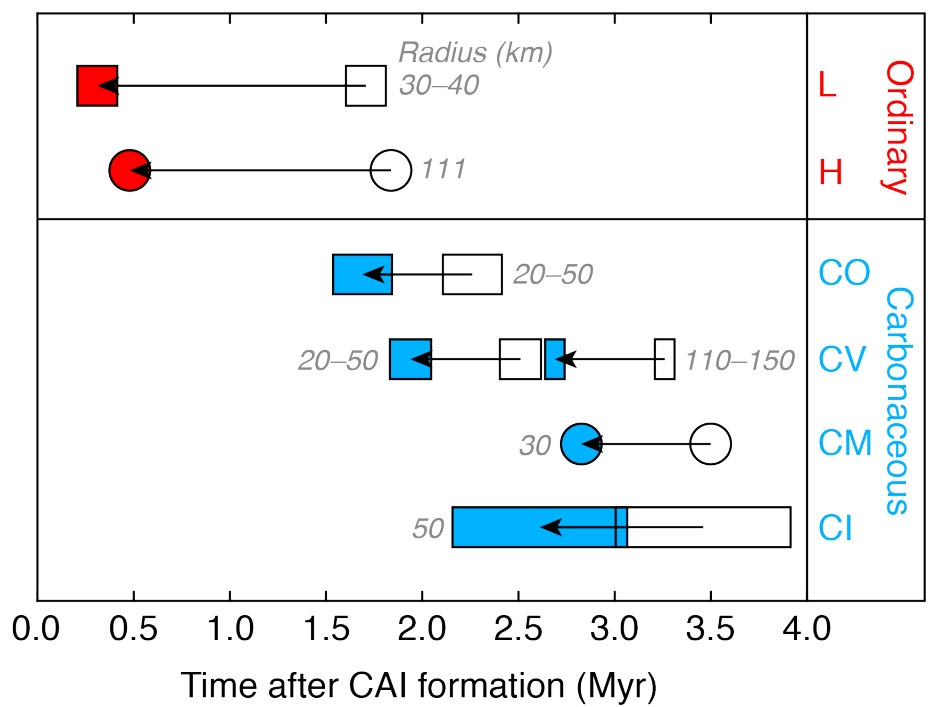

**Figure 3.** Accretion chronology of ordinary and carbonaceous chondrite parent bodies. The colored symbols represent the ages calibrated using the Al–Ti isotope correlation, whereas the open symbols represent the reported ages without the calibration. The numbers next to the symbols indicate assumed or inferred radii (km) of the parent bodies. The reported ages were from Doyle et al. (2015) for L, CO, and CV (20–50 km radius), Henke et al. (2012) for H, Jogo et al. (2017) for CV (110–150 km radius), Fujiya et al. (2012) for CM, and Fujiya et al. (2013) for CI chondrites.

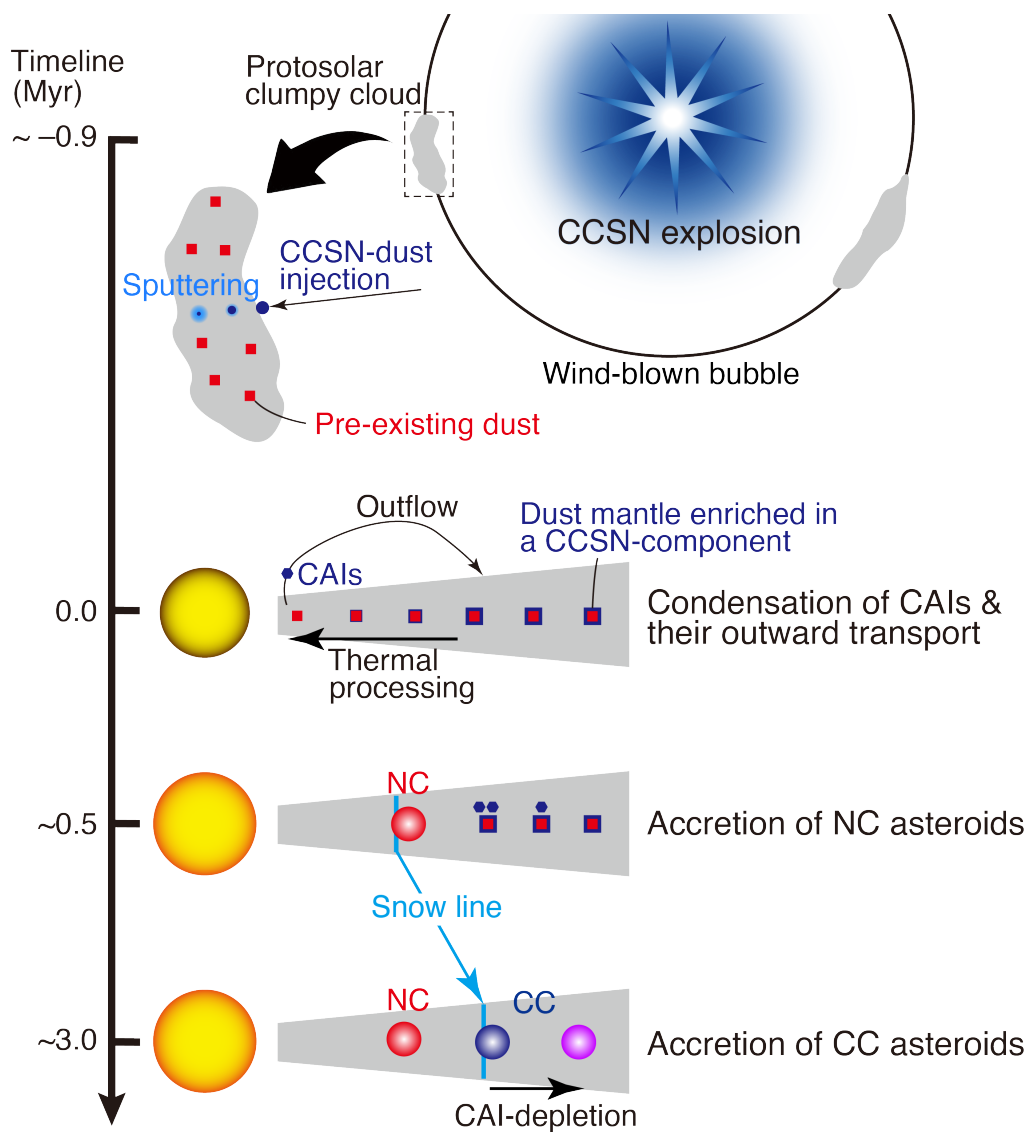

**Figure 4.** Schematic timeline of solar system formation. The ages shown are relative to the time of CAI formation. The Sun was born in association with a ~25 $M_\odot$ star. The protosolar clumpy cloud developed along the stellar wind-blown bubble was polluted by ejecta from the CCSN explosion at around −0.9 Myr through dust injection. The injected dust would be sputtered and release CCSN-derived nuclides including $^{26}$Al and Ti isotopes in the cloud, followed by condensation of these nuclides as mantles (blue) onto pre-existing dust of different stellar origins (red). Within the nascent inner disk around the protosun, these mantles would be preferentially destroyed by thermal processing, rendering the residual solids depleted in CCSN-derived nuclides. Besides, CAIs were condensed from the complementary gaseous reservoir rich in these nuclides and were transported to the outer disk by protostellar outflow. The parent asteroids of non-carbonaceous (NC) and carbonaceous chondrites (CC) were accreted in the inner and outer disks at ~0.5 and ~3.0 Myr, likely through the pile-up of inward-drifting pebbles around the snow line that has migrated outwards (Lichtenberg et al. 2021). Consequently, carbonaceous meteorites, in particular those bearing CAIs, are more enriched in CCSN-derived nuclides than the non-carbonaceous ones.

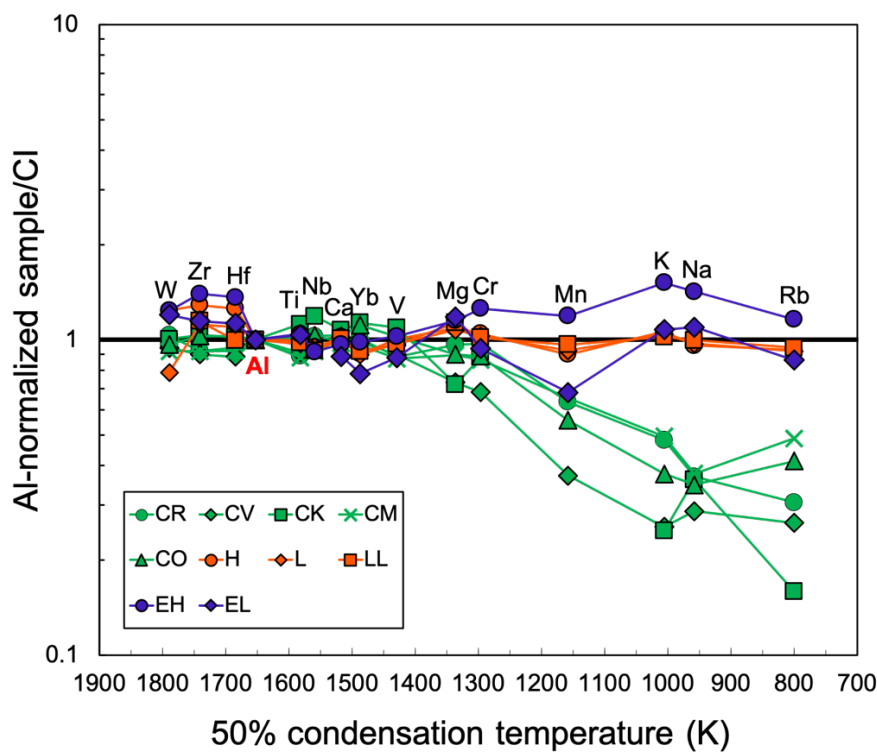

**Figure A1.** CI-normalized major and trace element abundances of chondrites. The data are plotted against the 50% condensation temperature of a solar composition gas at 10⁻⁴ bar (Lodders 2003). The data sources are Palme & O'Neill (2014) for CI, Lauretta et al. (2009) for CB, Bischoff et al. (1993) for CH, Braukmüller et al. (2018) for CR, CV, CK, CM, and CO, Wasson (1988) for H, L, and LL chondrites, and Wasson (1988) and Lee et al. (2000) for EH and EL chondrites.

**Table A1. Summary of Pb–Pb ages and Al–Ti isotope systematics of meteoritic samples**

| Sample | Pb–Pb age (Ma) | error | Ref. | $(^{26}Al/^{27}Al)_I$ | error | Ref. | $(^{26}Al/^{27}Al)_0$ | error | $\varepsilon^{50}Ti$ | error |
|---|---|---|---|---|---|---|---|---|---|---|
| CAIs | 4567.30 | 0.16 | 1, 2 | 5.25.E-05 | 1.90E-07 | 3 | 5.25E-05 | 7.98E-06 | 9.45 | 1.40 |
| D'Orbigny | 4563.51 | 0.18 | 4 | 3.98.E-07 | 1.50E-08 | 5 | 1.45E-05 | 2.55E-06 | -1.18 | 0.08 |
| Sahara 99555 | 4563.79 | 0.24 | 4 | 3.64.E-07 | 1.80E-08 | 5 | 1.02E-05 | 2.38E-06 | -1.18 | 0.08 |
| NWA 2976 | 4562.89 | 0.59 | 6 | 3.94.E-07 | 1.60E-08 | 6 | 2.59E-05 | 1.46E-05 | 3.09 | 0.11 |
| NWA 6704 | 4562.76 | 0.30 | 7 | 3.15.E-07 | 3.80E-08 | 8 | 2.35E-05 | 7.26E-06 | 2.10 | 0.40 |
| Asuka 881394 | 4564.95 | 0.53 | 9 | 1.48.E-06 | 1.20E-07 | 9 | 1.38E-05 | 7.03E-06 | -1.14 | 0.09 |

**Notes.** Errors are at the 2σ or 95% confidence interval.

**Table A2. Mass fractions of Ti and Al isotopes in the solar system and stellar ejecta**

|  | Ti-46 | Ti-47 | Ti-48 | Ti-49 | Ti-50 | Ref. |
|---|---|---|---|---|---|---|
| Solar system | 2.68E-07 | 2.47E-07 | 2.50E-06 | 1.87E-07 | 1.83E-07 | 1 |
| SNIa | 1.29E-05 | 5.53E-07 | 4.18E-04 | 1.51E-05 | 6.86E-09 | 2 |
| 15 $M_\odot$ CCSN | 1.03E-06 | 3.87E-07 | 9.44E-06 | 7.39E-07 | 3.39E-07 | 3 |
| AGB $s$-process | 1.63E-07 | 1.48E-07 | 1.48E-06 | 1.26E-07 | 1.36E-07 | 4 |
| He-core $s$-process in 30 $M_\odot$ star[a] | 3.21E-07 | 1.16E-07 | 2.32E-07 | 3.25E-07 | 9.38E-07 | 5 |
| C-shell $s$-process in 25 $M_\odot$ star[b] | 3.73E-08 | 2.90E-08 | 5.44E-08 | 5.30E-08 | 1.72E-07 | 6 |

|  | Al-26 | Al-27 |  |
|---|---|---|---|
| Solar system |  | 6.64E-05 | 1 |
| 25 $M_\odot$ CCSN* | 3.74E-06 |  | 7 |

**Notes.**

[a] The mass fractions are calculated assuming that the total mass of the ejecta is 28 $M_\odot$.

[b] The mass fractions are calculated assuming that the total mass of the ejecta is 23 $M_\odot$.

**References.** (1) Lodders 2003; (2) Maeda et al. 2010; (3) Rauscher et al. 2002; (4) Battino et al. 2019; (5) The et al. 2000; (6) Pignatari et al. 2010; (7) Limongi & Chieffi 2006.

**Table B1. Titanium isotope data for Asuka 881394**

| | $\varepsilon^{46}Ti$ | ± 2 s.e. | $\varepsilon^{48}Ti$ | ± 2 s.e. | $\varepsilon^{50}Ti$ | ± 2 s.e. |
|---|---|---|---|---|---|---|
| *Aliquot #1* | | | | | | |
| 1 | –0.18 | 0.10 | –0.03 | 0.07 | –1.20 | 0.12 |
| 2 | –0.15 | 0.16 | –0.09 | 0.07 | –1.16 | 0.16 |
| 3 | –0.19 | 0.13 | –0.07 | 0.05 | –1.13 | 0.09 |
| *Aliquot #2* | | | | | | |
| 1 | –0.18 | 0.17 | –0.07 | 0.07 | –1.16 | 0.15 |
| 2 | –0.23 | 0.15 | –0.04 | 0.06 | –1.07 | 0.16 |
| 3 | –0.18 | 0.13 | –0.07 | 0.06 | –1.12 | 0.16 |
| | | | | | | |
| Average ± 95% C.I. | –0.19 | 0.05 | –0.06 | 0.05 | –1.14 | 0.09 |